\def\di{\displaystyle}
\def\&{&\di}
\def\bg{\begin{eqnarray}\begin{array}{rcl}\displaystyle}
\def\bm#1{\begin{eqnarray}\begin{array}{#1}}
\def\eg{\end{array} &\di    &\di   \end{eqnarray}}
\def\bgo{\begin{eqnarray*}\begin{array}{rcl}\displaystyle}
\def\ego{\end{array} &\di    &\di \nonumber  \end{eqnarray*}}
\def\btensor#1#2{\renew\left#1\begin{array}{#2}\di}
\def\etensor#1{\end{array}\right#1}
\documentstyle[12pt]{article}
\def\MS{{{\hbox{MS}}}}

\def\LL{{\tiny{\hbox{LL}}}}
\def\pd{\partial}

\def\p{\phi}
\def\L{\Lambda}
\def\La{\L}
\def\m{\mu}
\def\l{\lambda}
\def\la{\l}

\def\h{\hbox{${1\over2}$}}
\def\q{\hbox{${1\over4}$}}
\def\ei{\hbox{${1\over8}$}}

\def\bib{\bibitem}

\def\ll{{\hbox{\tiny{LL}}}}
\def\ol{{\hbox{\tiny{$1$-loop}}}}
\def\tl{{\hbox{\tiny{$2$-loop}}}}
\def\tree{{\hbox{\tiny{tree}}}}

\def\be{\beta}

\def\ka{\kappa}

\def\e{\epsilon}
\def\d{\pd}
\def\k{\kappa}
\def\don{(\l\pd_{\l}+\h g^2\pd_{g^2})}

\def\dtwq{\q g^2\pd_{g^2}}

\def\beq{\begin{equation}}
\def\eeq{\end{equation}}
\def\bed{\begin{displaymath}}
\def\eed{\end{displaymath}}
\def\beqq{\begin{eqnarray}}
\def\eeqq{\end{eqnarray}}
\def\bedd{\begin{eqnarray*}}
\def\eedd{\end{eqnarray*}}

\def\rene{\renewcommand{\arraystretch}{1.6}}
\def\renew{\renewcommand{\arraystretch}{1}}
\rene
\begin{document}
\begin{flushright}
{\bf
FSUJ-TPI-20/96 \\
DIAS-STP-24/96\\
December 1996\\
}    
\end{flushright}

\par
\vskip .5 truecm
\large \centerline{\Large{\bf Multi-scale Renormalization}} 
\par
\vskip 1 truecm
\normalsize
\begin{center}
{\bf C.~Ford}\footnote{e--mail: cfo@tpi.uni-jena.de}\\
\sl{Theor.--Phys. Institut, Universit\"at Jena\\ 
Fr\"obelstieg 1\\
D--07743 Jena\\
Germany}
\vskip .5 truecm
{\bf C.~Wiesendanger}\footnote{e-mail: wie@stp.dias.ie. }\\
\sl{Dublin Institute for Advanced Studies\\
10 Burlington Road\\
Dublin 4\\
Ireland}
\end{center}
 \par
\vskip 2 truecm
\normalsize

\begin{abstract} 
{The standard $\MS$ renormalization
prescription is inadequate for dealing with multi-scale problems.
To illustrate this we consider the computation of the effective potential
in the Higgs-Yukawa model.
It is argued that  it is natural to employ a two-scale renormalization 
group. We give a modified version of a two-scale scheme introduced by 
Einhorn and Jones. In such schemes the beta functions necessarily 
contain potentially large logarithms of the RG scale ratios. For 
credible perturbation theory one must implement a large logarithms 
resummation on the beta functions themselves. We show how the 
integrability condition for the two RG equations allows one to 
perform this resummation.}
\end{abstract}
\clearpage
 
\begin{flushleft}{\sl Introduction}\end{flushleft}
\vspace{0.1cm}\noindent

Consider
the effective potential in the four-dimensional
Higgs-Yukawa model defined by the Lagrangian
\beq {\cal L}=\frac{1}{2}(\d_\m\p)^2-\frac{1}{2}m^2\p^2
-\frac{\l}{24}\p^4+\bar{\psi}i\d\!\!\!/\psi+g\bar{\psi}\p\psi+\L,
\eeq
where $\p$ is a real scalar field and $\psi=(\psi_1,...,\psi_N)^T$
are Dirac fields.
Here $\L$ is a ``cosmological constant'' term which enters non-trivially
into the renormalization group equation for the
effective potential \cite{kast}. It is well known how to perform
a loopwise perturbative expansion of the effective potential \cite{jac},
$V(\p)=V^{\tree}(\p)+\hbar V^{\ol}(\p)+\hbar^2V^{\tl}(\p)+...\,
$. Dimensional
regularization together with (modified) minimal subtraction gives
\beqq
V^{\tree}(\p)&=&\frac{\l}{24}\p^4+\frac{1}{2}m^2\p^2-\L,\\
V^{\ol}(\p)&=&\frac{(m^2+\h\l\p^2)^2}{4(4\pi)^2}\left[
\log\frac{m^2+\h\l\p^2}{\m^2}-\frac{3}{2}\right]
-\frac{ N 
g^4\p^4}{(4\pi)^2}\left[
\log \frac{g^2\p^2}{\m^2}-\frac{3}{2}\right]. \nonumber
\eeqq
 We have
a $\log \frac{m^2+\h\l\p^2}{\m^2}$ due to the ``Higgs'' loop
and a $\log \frac{g^2\p^2}{\m^2}$ associated with the fermionic
contribution to the one-loop potential. The two-loop potential
$V^{\tl}(\p)$ is quadratic in these logarithms, and in general the
$n$-loop potential is a $n$th order polynomial in the two logarithms.
Thus, for believable perturbation theory one must not only
have ``small'' couplings $\hbar \lambda$, $\hbar g^2$, but the two
logarithms must also be small. As was explained a long time ago
by Coleman and Weinberg (CW) \cite{cw}  one must make a ($\p$-dependent) choice
of $\m$ such that the logarithms are not too large. To relate the 
renormalized parameters at different scales one uses the renormalization group
(RG).
The CW procedure of RG ``improving'' the potential is equivalent to
a resummation of the large logarithms in the perturbation series.

However, it is not too difficult to see that if $m^2+\h\l\p^2\gg g^2\p^2$
(the heavy Higgs case) or $g^2\p^2\gg m^2+\h\l\p^2$ (ie. heavy 
fermions)
\sl there is no choice of $\m$ that will simultaneously render
both logarithms small \rm\,\cite{bando}. Thus, we are only able to
implement the CW
method when $m^2+\h\l\p^2\sim g^2\p^2$, ie. when we have essentially
a one-scale problem. In our opinion, the natural way to
deal
with this problem is to use a two-scale version of $\MS$. A multi-scale
version of $\MS$ was developed by Einhorn and Jones \cite{ej} (EJ).
An alternative multi-scale approach based on the Callan-Symanzik
equation has been outlined in refs. \cite{nish}.
However, the EJ scheme had two drawbacks. Firstly, although one has
several RG scales they do not ``track'' the relevant
logarithms
in an obvious way. Secondly, the beta functions contain logarithms of
the
RG scale ratios which render the perturbative beta functions useless
when these ratios are large.

In this letter we present a simple modification
of the EJ prescription where the RG scales naturally track the scales
which appear logarithmically in the effective potential. As in the
original EJ proposal one has potentially large logarithms in the
beta functions.
We argue that to deal with these logarithms one must implement a
large logarithms resummation on the beta functions themselves.
It is shown that the integrability condition for the two RG's allows
one to perform this resummation.

\begin{flushleft}{\sl The Einhorn-Jones Approach}\end{flushleft}
\vspace{.1cm}\noindent

Let us briefly recall the EJ multi-scale prescription.
To motivate their idea consider  the bare Lagrangian for
our Higgs-Yukawa problem written in terms of the usual
$\MS$ renormalized parameters.
\beqq 
{\cal L}_{\hbox{Bare}}&=&\frac{1}{2}Z_\p(\d_\m\p)^2-
\frac{1}{2}Z_\p Z_{m^2} m^2 \p^2-
\frac{1}{24}{\mu}^{\e}Z_\p^2 Z_\l  \l \p^4\nonumber\\
&&
+
Z_\psi\bar{\psi}i\d\!\!\!/\psi+
Z_\psi Z_{g^2}^{\h} {Z_\p}^{\h} {\m}^{\h \e} g\bar{\psi}\p\psi+
\L+{\mu}^{-\e}(Z_\L-1)m^4\l^{-1}, \label{3}
\eeqq
where $\e=4-d$ is the dimensional continuation parameter,
and all the $Z_.$ factors have the form
$Z_.=1+\hbox{\it pole terms only}$. Notice that the $\MS$ RG scale
$\mu$ enters eqn. (\ref{3}) in three places.
The EJ idea was simply to replace the three occurrences of
$\m$ in (\ref{3})  with three \sl independent \rm\,  RG scales
$\k_1$, $\ka_2$, $\ka_3$, so that
\bm{rclrclrcl}
\l_{B}&=&\k_1^{\e}Z_\l \l,&
g^2_{B}&=&\k_2^{ \e} Z_{g^2} g^2,&
\L_{B}&=&\La+\k_3^{-\e}(Z_\L-1) m^4\l^{-1},\\
m^2_B&=&Z_{m^2}m^2,&
\p_B&=&Z_{\p}^{\h}\p,&
\psi_B&=&Z_{\psi}^{\h}\psi.
\eg
As in standard $\MS$ the $Z_.$ factors are defined
by the requirement that the effective action is finite
when written in terms of the renormalized parameters
and the restriction that the $Z_.$ factors have
the form $Z_.=1+\hbox{\it pole terms only}$. Note that the $Z_.$
factors will \sl not \rm be the same as the $\MS$ $Z_.$ factors
(except where $\k_1=\k_2=\k_3$). In the EJ scheme the $Z_.$'s contain
logarithms of the RG scale ratios.

We now have three separate RG equations associated with the independent
variations of the three RG scales. We also have three sets of beta
functions 
\beq \label{5} {}_i\beta_\la=\k_i \frac{d}{d\k_i}\la;\quad
i=1,2,3 \eeq
and similarly for the other parameters. 
It is straightforward to compute the one-loop RG functions in the EJ
scheme. For example, the one-loop beta functions for the two
couplings are
\beqq\label{6}
{}_2\be_{g^2}^\ol&\!\!\!=\!\!\!&\frac{(4N+6)\hbar g^4}{(4\pi)^2},\quad\!\!\!\!
{}_1\be_{g^2}^\ol={}_3\beta_{g^2}^\ol=0,\nonumber \\
{}_1\be_\l^\ol&\!\!\!=\!\!\!&\frac{\hbar}{(4\pi)^2}(3\l^2+48Ng^4),\quad\!\!\!\!
{}_2\be_\l^\ol=\frac{\hbar}{(4\pi)^2}(8N\l g^2-96 Ng^4),\quad\!\!\!\!
{}_3\be_\l^\ol=0.
\eeqq
However, if one were to compute
the two-loop beta functions, one would find terms proportional
to $\log \frac{\k_1}{\k_2}$, and in general the $n$-loop RG functions
contain $\log^{n-1}\frac{\k_1}{\k_2}$ terms (as well as lower powers of
the logarithm). Therefore, unlike in standard $\MS$ we \sl
cannot trust the perturbative RG functions. \rm

In the EJ scheme the one-loop potential is 
\beqq
V^\ol&=&\frac{(m^2+\h\l\p^2)^2}{4(4\pi)^2}
\left(\log\frac{m^2+\h\l\p^2}{\k_1^2}-\frac{3}{2}\right)\nonumber\\
&&-\frac{Ng^4\p^4}{(4\pi)^2}\left(
\log\frac{g^2\p^2}{\k_2^2}+\log\frac{\k_1^2}{\k_2^2}-\frac{3}{2}
\right)
+\frac{m^4}{2(4\pi)^2}\log\frac{\k_1}{\k_3}
\eeqq
whereas what we want is a \sl two-scale \rm\, scheme where the one-loop potential
is
\beq
V^\ol=\frac{(m^2+\h\l\p^2)^2}{4(4\pi)^2}\left(
\log\frac{m^2+\h\l\p^2}{\k_1^2}-\frac{3}{2}\right)-
\frac{Ng^4\p^4}{(4\pi)^2}\left(
\log\frac{g^2\p^2}{\k_2^2}-\frac{3}{2}\right).\label{pot}
\eeq

\begin{flushleft}{\sl Attaching RG scales to the kinetic terms}
\end{flushleft}
\vspace{.1cm}\noindent

The reason for this mismatch is that in the EJ scheme
the RG scales are attached to the coupling constant terms in the 
Lagrangian, but the arguments of the logarithms in the effective potential
are associated with the kinetic part of the action. However,
there is no reason  why independent
RG scales should not be ``attached'' to the kinetic terms. That is,
instead of (\ref{3}) we propose  to start with
\beqq
{\cal L}_{\hbox{Bare}}&=&{1\over2}{\k_1}^{-\h\epsilon}
Z_\p(\partial_\m\p)^2-{1\over2}Z_\p Z_{m^2}m^2\p^2-\frac{Z_\l {Z_\p}^2}{24}
\l\p^4\nonumber\\
&&+{\k_2}^{-\q\epsilon}Z_\psi\bar\psi i\partial\!\!\!/ \psi
+Z_{g^2}^{\h} Z_{\psi} Z_\p^{\h}g\bar\psi\p\psi
+\Lambda+(Z_\Lambda -1)m^4\l^{-1}.\label{bare2}
\eeqq
As before we insist that the \sl renormalized \rm\, couplings
$\l$ and $g$ are dimensionless. However, by altering the dimensions
of the \sl renormalized dimensionful \rm\, objects $m^2$, $\phi$ and $\psi$ we can
reshuffle the RG scales into the kinetic terms only.
Again we require that the $Z_.$ factors are of the form
$Z_.=1+\hbox{\it pole terms only}$. 

From (\ref{bare2}) one can read off the relation between the
bare and renormalized parameters
\beqq
\l_B&=&{\k_1}^{ \epsilon}Z_\l\l,\quad
g^2_B={\k_1}^{\h\epsilon}{\k_2}^{\h\epsilon}Z_{g^2} g^2,\quad
m^2_B=\k_1^{\h\e}Z_{m^2}m^2,\nonumber \\
\p_B&=&{\k_1}^{-\q \e}Z_{\p}^{\h}\p,\quad
\psi_B={\k_2}^{-\ei \e}Z_{\psi}^{\h}\psi,\quad
\L_B=\L+(Z_\L -1)m^4\l^{-1}.
\eeqq
In this ``modified'' EJ approach
the one-loop effective potential is indeed given  by (\ref{pot}).
It is straightforward to obtain the beta functions in terms of
the $Z_.$ factors. As in standard MS one only needs the simple
pole terms in the $Z_.$'s
\renew\bm{rclcrcl}
{}_1\be_{\l}& \!\!\! =& \!\!\! -\e\l+\l\don z_{\l}
&&{}_1\be_{g^2} & \!\!\! =& \!\!\! -\h\e g^2+g^2\don z_{g^2}\\
{}_2\be_{\l} & \!\!\! =& \!\!\! \h\l g^2\pd_{g^2} z_{\l}&&
{}_2\be_{g^2} & \!\!\! = & \!\!\! 
-\h\e g^2+\h g^4\pd_{g^2}z_{g^2},\vspace{2mm}\\
{}_1\be_{m^2}& \!\!\! =& \!\!\! -\h\e m^2+m^2\don z_{m^2}&&
{}_1\be_{\Lambda}& \!\!\! =& \!\!\! {\lambda}^{-1}m^4\left(
\l \partial_{\lambda}+\h g^2\partial_{g^2}\right)z_{\Lambda}
\\
{}_2\be_{m^2}& \!\!\! =& \!\!\! \h m^2 g^2\partial_{g^2}z_{m^2}&&
{}_2\be_{\Lambda}& \!\!\! =& \!\!\! \h{\lambda}^{-1}m^4g^2\partial_{g^2}z_{\Lambda}
\vspace{2mm}\\
{}_1\be_{\phi}& \!\!\! =& \!\!\! \q \e +\h\don z_{\p}&&
{}_1\be_{\psi}& \!\!\! =& \!\!\! \h\don z_{\psi}\\
{}_2\be_{\phi}& \!\!\! =& \!\!\!  \dtwq z_{\p}&&
{}_2\be_{\psi}& \!\!\! =& \!\!\! \ei \epsilon+\q g^2\partial_{g^2}z_{\psi}.
\eg\rene
Here $z_.$ denotes the coefficient of $1/\e$ in
$Z_.$. To compute the one-loop beta functions one needs the
one-loop contribution to the $z_.$'s which are identical
to the MS $z_.$'s. However, at higher loops the $z_.$'s
(and hence the beta-functions) contain logarithms of the RG scale ratios.
When these logarithms are large the one-loop multi-scale RG functions
cannot be trusted. To get reliable approximations to the beta functions
we must implement a large logarithms resummation on the beta functions
themselves.

\begin{flushleft}{\sl Integrability of the two RG equations}\end{flushleft}
\vspace{.1cm}\noindent

The effective potential satisfies two RG equations associated with the
independent variations of the two RG scales
\beq {\cal D}_i V=0;\quad i=1,2,
\eeq
where
\beq
{\cal D}_i=\k_i\frac{\pd}{\pd\k_i}+
{}_i\be_{g^2}\frac{\pd}{\pd g^2}+
{}_i\be_\l \frac{\pd}{\pd \l}+
{}_i\be_{m^2}\frac{\pd}{\pd m^2}+
{}_i\be_{\L}\frac{\pd}{\pd \L}+
{}_i\be_{\p}\p\frac{\pd}{\pd \p}.
\eeq
The integrability condition for these two equations is simply
\begin{equation}[{\cal D}_1,{\cal D}_2]=0,\label{int}\end{equation}
ie. independent variations of the two RG scales commute.
Recently, it has been argued that in any multi-scale scheme the
integrability
condition imposes a strong constraint on the form of the multi-scale
RG functions \cite{fowi}. Here we show  that for EJ type schemes
this constraint is strong enough
to implement a large logs resummation on the beta functions.

If we now insert the expressions
 for the multi-scale beta
functions in terms of the
$z_.$'s into the integrability condition (\ref{int}) we obtain
a set of (non-linear) partial differential
equations for the $z_.$'s.
\beqq \label{hard}
0&=&(4\pi)^{-2}\hbar\pd_t\left(g^2\pd_{g^2}+\l \pd_{\l}\right)z_\l
+\h\l^2g^2(\pd_\l z_\l)\pd_{g^2}\pd_\l z_\l
-\h\l g^2 (\pd_{g^2}z_\l)\pd_\l z_\l\nonumber\\
&&-\h g^2\l^2(\pd_{g^2}z_\l){\pd_\l}^2 z_\l
+\h g^2\l (\pd_\l z_{g^2})\pd_{g^2}z_{\l}
+\h g^4\l (\pd_\l z_{g^2}){\pd_{g^2}}^2z_\l\nonumber\\
&&-\h g^4\l (\pd_{g^2} z_{g^2})\pd_\l \pd_{g^2} z_\l,\nonumber\\
0&=&(4\pi)^{-2}\hbar\pd_t\left(g^2\pd_{g^2}+\l\pd_{\l}\right)z_{g^2}
+\h\l g^2(\pd_\l z_{g^2})\pd_{g^2} z_{g^2}+
\h\l g^4(\pd_\l z_{g^2}) \pd_{g^2}^2 z_{g^2}\nonumber\\
&&-\h\l g^4 (\pd_{g^2}z_{g^2})\pd_\l \pd_{g^2}z_{g^2}
+\h\l^2 g^2(\pd_\l z_\l) \pd_\l \pd_{g^2}z_{g^2}
-\h\l g^2 (\pd_{g^2} z_\l)\pd_\l z_{g^2}\\
&&-\h\l^2 g^2 (\pd_{g^2}z_\l)\pd_\l^2 z_{g^2},\nonumber\\
0&=&(4\pi)^{-2}\hbar\pd_t\left(g^2\pd_{g^2}+\l\pd_{\l}\right)
z_{m^2}+\h\l^2g^2(\pd_\l z_\l) \pd_{g^2}z_{m^2}-
\h g^2(\pd_{g^2}z_\l)\pd_\l z_{m^2}\nonumber\\
&&-\h\l g^2 (\pd_\l z_{\l})\pd_\l^2 z_{m^2}
+\h\l g^2(\pd_\l z_{g^2})\pd_{g^2}z_{m^2}+
\h\l g^4(\pd_{\l} z_{g^2})\pd_{g^2}^2z_{m^2}\nonumber\\
&&-\h\l g^4(\pd_{g^2}z_{g^2})\pd_\l \pd_{g^2} z_{m^2},\nonumber
\eeqq
where $(4\pi)^2t=\hbar \log (\k_1/\k_2)$. One can obtain similar
equations for $z_\p$, $z_\L$ and $z_\psi$.
These equations 
\sl which are exact to all orders in perturbation theory, \rm
together with the boundary conditions 
that $z_.$ collapses to the MS result at $\k_1=\k_2$,
fully determine all the logarithmic terms in the $z_.$'s (and hence
the beta functions). To sum the \sl leading logarithms \rm (LL) 
in the $z_.$'s
we must solve equations (\ref{hard})
using the \sl one-loop \rm MS $z_.$'s as boundary
conditions at $t=0$ (ie. $\k_1=\k_2$)
\bm{rclrcl}\label{onelz}
(4\pi)^2z_{\l}^\ol&=&{\hbar}(3\l +8Ng^2-48Ng^4\l^{-1}), &
(4\pi)^2z_{g^2}^\ol&=&(4N+6)\hbar g^2,\\
(4\pi)^2z_{m^2}^\ol&=&\hbar m^2\left(\l+4Ng^2\right),&
(4\pi)^2z_{\Lambda}^\ol&=&\h\hbar \l m^4,\\
(4\pi)^2z_\p^\ol&=&-2N\hbar g^2,&
(4\pi)^2
  z_\psi^\ol&=&\hbar g^2\\
\eg
In general we do not know how to solve equations of the type
(\ref{hard}). However, for the LL
 calculation the following ansatz
is applicable
\beq
(4\pi)^2 z_\l^\ll=\hbar[
\l a(s) +g^2 b(s) +g^4\l^{-1} c(s)],\label{ansatz}
\eeq
where $s=g^2 t=\hbar g^2 (4\pi)^{-2}\log (\k_1/\k_2)$,
and $z_{g^2}^\ll$, $z_{\phi}^\ll$, $z_{\psi}^\ll$  are \sl identical \rm\,
to the one-loop results (\ref{onelz})
(ie. the
RG functions for $g^2$, $\p$ and $\psi$ contain no leading logarithms).
The functions
$a(s)$, $b(s)$, $c(s)$ satisfy the following (coupled) ordinary
differential equations
\beqq
sa''(s)+2a'(s)&=&\h a(s)\bigl( b(s)+s
b'(s)\bigr)+(2N+3)sa'(s),\nonumber\\
sb''(s)+2b'(s)&=&sa'(s)c(s)+\bigl(2c(s)+sc'(s)\bigr)a(s),\nonumber\\
sc''(s)+2c'(s)&=&\h c(s)\bigl(b(s)+sb'(s)\bigr)-(2N+3)\bigl(
2c(s)+sc'(s)\bigr),\label{odes}
\eeqq
where $'=d/ds$, and the initial values are
$a(0)=3$, $b(0)=8N$, $c(0)=-48N$.
The corresponding RG functions are
\bg
\begin{array}{rclcrclcrcl}
(4\pi)^2{}_1\beta_{g^2}^\ll&=&(2N+3)\hbar g^4,&
(4\pi)^2{}_1\beta_{\phi}^\ll&=&-N\hbar g^2,&
(4\pi)^2{}_1\beta_{\psi}^\ll&=&\h\hbar g^2,\\
(4\pi)^2{}_2\beta_{g^2}^\ll&=&(2N+3)\hbar g^4,&
(4\pi)^2{}_2\beta_{\phi}^\ll&=&-N\hbar g^2,&
(4\pi)^2{}_2\beta_{\psi}^\ll&=&\h\hbar g^2,
\end{array}\\
\begin{array}{rcl}
(4\pi)^2{}_1\beta_\l^\ll&=&\hbar\l^2\bigl(a(s)+\h sa'(s)\bigr)+\h\hbar\l g^2\bigl(
b(s)+sb'(s)\bigr)+\h\hbar g^4sc'(s),\\
(4\pi)^2{}_2\beta_\l^\ll&=&\h\hbar\l^2sa'(s)+\h\hbar\l g^2\bigl(
b(s)+sb'(s)\bigr)+\h\hbar g^4\bigl(2c(s)+sc'(s)\bigr).
\end{array}
\eg

We have seen that 
$z_{g^2}^\ll$, $z_{\phi}^\ll$, and $z_{\psi}^\ll$ are trivial, while
$z_\l^\ll$ has the form (\ref{ansatz}).
However, the problem of computing
$z_{m^2}^\ll$ and $z_\L^\ll$ is more tricky. They satisfy the
following
partial differential equations
\bm{rcl}
0&=& g^2\left(s\pd_s^2+2\pd_s\right)z_{m^2}^\ll+
\h g^2\left(\l^2a(s)-g^4c(s)\right)\pd_\l\pd_{g^2}z_{m^2}^\ll
-(2N+3)\l g^4\pd_\l\pd_{g^2}z_{m^2}^\ll \\
&&-\left[
\l^2a'(s)s+\l g^2\left(b(s)+b'(s)s\right)+g^4\left(2c(s)+c'(s)s
\right)\right](\pd_\l+\l\pd_\l^2)z_{m^2}^\ll \\
0&=&\l^{-1}g^2\left(s\pd_s^2+2\partial_s\right)z_{\Lambda}^\ll
+\left(\l^2 a(s)-g^4 c(s)\right)
\left[\h\l^{-1}g^2\pd_\l\pd_{g^2}z_{\Lambda}^\ll-\h\l^{-2}g^2\pd_{g^2}
z_{\Lambda}^\ll \right]\\
&&-\left[
\l^2a'(s)s+\l g^2\left(b(s)+b'(s)s\right)+g^4\left(2c(s)+c'(s)s
\right)\right]\pd_\l^2z_{\Lambda}^\ll \\
&&-(2N+3)\l g^4\pd_\l\pd_{g^2}z_{\Lambda}^\ll
+g^2(\pd_\l z_{m^2}^\ll)\pd_{g^2}z_{\Lambda}^\ll-g^2(\pd_{g^2}z_{m^2}^\ll)
\pd_{\lambda}z_{\Lambda}^\ll. \label{pdes}
\eg
Although we do not have closed forms for $z_\l^\ll$,
$z_{m^2}^\ll$ and $z_{\Lambda}^\ll$ in the LL approximation, it is easy to
check
that for the special cases $N=0$ and $N\rightarrow\infty$ there are no
logarithms  in the $z_.$'s. This is what we would expect since here we
are
dealing with single scale problems.

\begin{flushleft}{\sl Discussion}\end{flushleft}
\vspace{.1cm}\noindent

We have presented a modification of the EJ multiscale prescription
where the RG scales naturally track the logarithms present in the
perturbation series for the effective potential. This involves the
attachment
of independent RG scales to the kinetic terms in the action rather
than to
the interaction terms. We have illustrated our proposal using
a simple two-scale Higgs-Yukawa model. It is straightforward to
apply this approach to other  renormalizable theories
\footnote{  The situation is a little more
complicated
in gauge theories where one has interactions involving derivatives.
Moreover, one should ensure that the attachment of independent 
RG scales does not spoil gauge invariance.}.
 
As in the EJ prescription there are  potentially large logarithms in the
beta functions. We have shown that the integrability condition for the two
RG equations is sufficient (in principle) to implement a large logarithms
resummation on the beta functions. However we have seen that even
in the simple Higgs-Yukawa model, the calculation of the LL beta
functions
is quite involved (one needs to solve the system of ODE's (\ref{odes})
to determine the ${}_i\beta_{\l}^\LL$, while ${}_i\beta_{m^2}^\LL$
and ${}_i\beta_{\Lambda}^\LL$ require a solution of eqns. (\ref{pdes})).
An  analytic or numerical solution of these equations would be helpful.

\section*{Acknowledgements}
We are grateful to D. O' Connor, J.M. Pawlowski and C.R. Stephens
for stimulating discussions. C. F. thanks DIAS for
warm hospitality and financial support. C. W. has been
partially supported by Schweizerischer Nationalfonds.

\end{document}